\documentstyle[12pt]{article}
\begin{document}
\def\be{\begin{equation}}
\def\ee{\end{equation}}
\title{\bf {Bouncing cosmological solutions and their stability }}
\author{M. R. Setare, \footnote{E-mail: rezakord@ipm.ir},
\\{Physics Department, Payame Noor University. Bijar. Iran}\\F. Darabi \footnote{E-mail:
f.darabi@azaruniv.edu}\\ {Research Institute for Astronomy and
Astrophysics of Maragha
 }\\{P. O. Box 55134-441, Maragha, IRAN} }
\date{\small{}}
\maketitle
\begin{abstract}
In the present paper we consider the bouncing braneworld scenario,
in which the bulk is given by a five-dimensional charged AdS black
hole spacetime with matter field confined in a $D_3$ brane. Then,
we study the stability of solutions with respect to homogeneous
and isotropic perturbations. Specifically, the AdS black hole with
zero ADM mass and charge, and open horizon is an attractor, while
the charged AdS black hole with zero ADM mass and flat horizon, is
a repeller.
\end{abstract}
\newpage

 \section{Introduction}

Motivated by string/M theory, the AdS/CFT correspondence, and the
hierarchy problem of particle physics, braneworld models were
studied actively in recent years \cite{Hora96}-\cite{Rand99}. In
these models, our universe is realized as a boundary of a higher
dimensional spacetime. In particular, a well studied example is
when the bulk is an AdS space. In the cosmological context,
embedding of a four dimensional Friedmann-Robertson-Walker
universe was also
considered when the bulk is described by AdS or AdS black hole \cite%
{Nihe99,AdSbhworld}. In the latter case, the mass of the black hole
was found to effectively act as an energy density on the brane with
the same equation of state of radiation. Representing radiation as
conformal matter and exploiting AdS/CFT correspondence, the
Cardy-Verlinde formula \cite{Verl00} for the entropy was found for
the universe ( see \cite{Seta02}, for the entropy formula in
the case of dS black hole ).\\
In either of the above cases, however, the cosmological evolution
on the brane is modified at small scales. In particular, if the
bulk space is taken to be an AdS black hole {\it with charge}, the
universe can `bounce' \cite{PEL}. That is, the brane makes a
smooth transition from a contracting phase to an expanding phase.
From a four-dimensional point of view, singularity theorems
\cite{hawk} suggest that such a bounce cannot occur as long as
certain energy conditions apply. Hence, a key ingredient in
producing the bounce is the fact that the bulk geometry may
contribute a negative energy density to the effective
stress-energy on the brane \cite{negative}. At first sight these
bouncing braneworlds are quite remarkable, since they provide a
context in which the evolution evades any cosmological
singularities while the dynamics is still controlled by a simple
(orthodox) effective action. In particular, it seems that one can
perform reliable calculations without deliberation on the effects
of quantum gravity or the details of the ultimate underlying
theory. Hence, several authors \cite{others,higherd,kanti,mset}
have pursued further developments for these bouncing braneworlds.
However the authors of \cite{mh} have found that generically these
cosmologies are in fact singular. In particular, they  have shown
that a bouncing brane must cross the Cauchy horizon in the bulk
space. However, the latter surface is unstable when arbitrarily
small excitations are introduced in the bulk spacetime.\\
In this paper we describe solutions of the bouncing braneworld
theory and also determine their stability. To do this, we use a
set of convenient phase-space variables similar to those
introduced in \cite{CS,GE}. The critical points of the system of
differential equations in the space of these variables describe
interesting non-static solutions. A method for evaluating the
eigenvalues of the critical points of the Friedmann and Bianchi
models was introduced by Goliath and Ellis \cite{GE} and further
used in the analysis by Campos and Sopuerta \cite{CS} of the
Randall Sundrum braneworld theory. These latter authors gave a
complete description of stationary points in an appropriately
chosen phase space of the cosmological setup and investigated
their stability with respect to homogeneous and isotropic
perturbations. The authors worked in the frames of the Randall
Sundrum braneworld theory without the scalar-curvature term in the
action for the brane. Cosmological solutions and their stability
with respect to homogeneous and isotropic perturbations in the
braneworld model with the scalar-curvature term in the action for
the brane have further been studied by Iakubovskyi and Shtanov
\cite{shta}.

\section{ $D_3$ Brane with Radiative Matter}

We start by considering a $4-$dimensional brane in a space-time
described by a $5-$dimensional charged AdS black hole. The
background metric is then given by \cite{CEJM}
\begin{equation}
ds_{5}^2 = - h(a) dt^2 + {{da^2}\over{h(a)}} + a^2
d\Omega_{k,3}^2, \label{metric}
\end{equation}
where
\begin{equation}
h(a) = k - {\omega_{4} M\over{a^{2}}} + {3 \omega_{4}^2 Q^2 \over
{16 a^{4}}} + {a^2\over{L^2}}, \label{component}
\end{equation}
in which $M$ and $Q$ are ADM mass and charge, respectively, $L$ is
the curvature radius of  space, $k$  is a discrete parameter which
indicates the horizon topology, namely $k=$  +1, 0 and -1 for a
spherical, flat and hyperbolic geometry respectively, and
\begin{equation}
\omega_{4} =\frac{16 \pi  G_5}{3 V_3}.
 \label{omega}
\end{equation}
Here $G_5$ is the five-dimensional Newton constant, and  $V_3$ is
the dimensionless volume element associated with $d\Omega^2_{k,3}$
(a three-dimensional spacelike  hypersurface of constant
curvature). \\
Tuning the brane cosmological constant to zero, we get the
following induced equation on the brane:
\begin{equation}
H^2 = -{k\over {a^2}} + {\omega_{4} M\over {a^{4}}} - { 3
\omega_{4}^2 Q^2\over{16 a^{6}}}, \label{freedman}
\end{equation}
where $H = {\dot a\over a}$ is the Hubble parameter and dot
denotes derivation with respect to the brane time $\tau$. If we
consider the more physically relevant case in which a perfect
fluid with equation of state of radiation is present on the
brane, then the first Friedmann equation takes the following form
\begin{equation}
 H^{2}=\frac{-k}{a^{2}}+\frac{8\pi
G_{4}}{3}\rho-\frac{1}{l^{2}}+\frac{4\pi}{3M_{p}^{2}\rho_{0}}(\rho_{0}+\rho_{br})^{2}-
{ 3 \omega_{4}^2 Q^2\over{16 a^{6}}}, \label{shab1}
 \end{equation}
where $\rho$ is the energy density defined by $\rho=E_{4}/V$ and
the four-dimensional energy $E_4$ is given by
 \be E_4=\frac{-3V_3}{16\pi G_5 a}M=\frac{l}{r}E, \label{ener2} \ee where $E$ is the
 thermodynamical energy of the black hole. Also in Eq.(\ref{shab1}) $\rho_{0}$ is the tension of the brane,
while $\rho_{br}$ is the energy density of the radiation. The
Hubble equation can be rewritten as \be
H^{2}=\frac{-k}{a^{2}}+\frac{8\pi
G_{4}}{3}\rho+\frac{\Lambda_{4}}{3}+\frac{8\pi}{3M_{p}^{2}}(\frac{\rho_{br}^{2}}{2\rho_{0}}+\rho_{br})-{
3 \omega_{4}^2 Q^2\over{16 a^{6}}}, \label{shab2} \ee where \be
\Lambda_{4}=\frac{4\pi
\rho_{0}}{M_{p}^{2}}-\frac{3}{l^{2}}=\Lambda_{br}-\frac{3}{l^{2}},
\label{cosmo} \ee is the effective cosmological constant of the
brane. By tuning the bulk cosmological constant and the brane
tension $\Lambda_{br}$, the effective four dimensional
cosmological constant $\Lambda_{4}$ can be set to zero. This
critical brane is what we would like to consider in the following.
The radiation energy density $\rho_{br}$ on the brane is  \be
\rho_{br}=\frac{\rho_{r}}{a^{4}}, \label{tens} \ee where
$\rho_{r}$ is a constant, hence we can rewrite the cosmological
equation (\ref{shab2}) as \be
H^{2}=\frac{-k}{a^{2}}+(\omega_4M+\frac{8\pi
\rho_{r}}{3M_{p}^{2}})\frac{1}{a^4}-\frac{3\omega_{4}^{2}Q^2}{16a^6}+\frac{4\pi
\rho_{r}^{2} }{3M_{p}^{2}\rho_{0}a^{8}}. \label{modhub2}\ee By
defining \be A=(\omega_4M+\frac{8\pi \rho_{r}}{3M_{p}^{2}})
\label{eqa}, \ee \be B= \frac{3\omega_{4}^{2}Q^2}{16}
\label{eqb},\ee and \be C=\frac{4\pi \rho_{r}^{2}
}{3M_{p}^{2}\rho_{0}},\label{eqc} \ee Eq.(\ref{modhub2}) takes on
the following form in terms of the above parameters
\be
H^2=\frac{-k}{a^2}+\frac{A}{a^4}-\frac{B}{a^6}+\frac{C}{a^8}
\label{par}\ee

\section{Stability of the Bouncing Solutions}

In this section, we describe solutions of the braneworld theory
under investigation and also determine their stability. To do
this, we use a set of convenient phase-space variables similar to
those introduced in \cite{CS,GE}. Now, we introduce the notation
similar to those of \cite{CS}
\begin{equation}\label{omeq}
\Omega_{k}=\frac{-k}{a^2H^2}=\frac{-k}{\dot{a}^{2}}, \hspace{1cm}
\Omega_{A}=\frac{A}{a^4H^2},\hspace{1cm}
\Omega_{B}=\frac{-B}{a^6H^2},\hspace{1cm}\Omega_{C}=\frac{C}{a^8H^2}.
\end{equation}
and work in the $4$-dimensional $\Omega$-space
$(\Omega_{k},\Omega_{A},\Omega_{B},\Omega_{C})$. In this space,
the $\Omega$ parameters are not independent since the Friedmann
equation (\ref{freedman}) reads
\begin{equation}\label{constraint}
\Omega_{k}+\Omega_{A}+\Omega_{B}+\Omega_{C}=1.
\end{equation}
In this case the state space defined by the variables
$(\Omega_{k},\Omega_{A},\Omega_{B},\Omega_{C})$ is no longer
compact ( because now $\Omega_{B}< 0$ ). However, we can introduce
another set of variables describing a compact state space.
Firstly, instead of using the Hubble function $H$ we will use the
following quantity
\begin{equation}
D=\sqrt{H^2+{B\over{a^{6}}}}, \label{ppp}
\end{equation}
and from it, let us define the following dimensionless
variables
\begin{equation}
Z=\frac{H}{D},\label{Z}
\end{equation}
\begin{equation}
\tilde{\Omega}_{k}=\frac{-k}{a^2D^2},\label{k}
\end{equation}
\begin{equation}
\tilde{\Omega}_{A}=\frac{A}{a^4D^2},\label{A}
\end{equation}
\begin{equation}
\tilde{\Omega}_{C}= \frac{C}{a^8D^2}\label{C}.
\end{equation}

From these definitions we see that now the case $H = 0$ is
included. Moreover, the Friedmann equation takes the following
form
\begin{equation}
\tilde{\Omega}_{k}+\tilde{\Omega}_{A}+\tilde{\Omega}_{C}
=1,\label{const1}
\end{equation}
which, together with the fact that $-1\leq Z \leq 1$ [see Equation
(\ref{ppp})], implies that the state space defined by the new
variables is indeed compact.

Introducing the primed time derivative
\begin{equation}
' = \frac{1}{D} \frac{d}{dt} \, ,
\end{equation} one obtains the
system of first-order differential equations
\begin{equation}\begin{array}{l}\label{omemgaeq}
D'=-ZD[1+(q-2\Omega_B)Z^2]\, , \smallskip \\
Z'=-Z^2[q-(q-2\Omega_B)Z^2]\, , \smallskip \\
\tilde{\Omega}_{k}'=2\tilde{\Omega}_{k}(q-2\Omega_B)Z^3\, , \smallskip \\
\tilde{\Omega}_{A}'=2\tilde{\Omega}_{A}[-1+(q-2\Omega_B)Z^2]Z\, , \smallskip \\
\tilde{\Omega}_{C}' = 2 \tilde{\Omega}_{C} [-3+(q-2\Omega_B)Z^2]Z\,
,
\end{array}
\end{equation}
where
\begin{equation}\label{qeq}
1+(q-2)Z^2=4\tilde{\Omega}_{C}-\tilde{\Omega}_{k}.
\end{equation}
The evolution equation for $D$ is not coupled to the rest, so we
will not consider it for the dynamical study. Therefore, we just
study the dynamical system for the variables
$\tilde{\Omega}\equiv(Z, \tilde{\Omega}_{k}, \tilde{\Omega}_{A},
\tilde{\Omega}_{C})$, determined by the equations (\ref{omemgaeq}).

The behavior of this system of equations in the neighborhood of its
stationary point is determined by the corresponding matrix of its
linearization. The real parts of its eigenvalues tell us whether the
corresponding cosmological solution is stable or unstable with
respect to the homogeneous perturbations \cite{shta}.
 To begin with, we have to find
the critical points of this dynamical system, which can be
written in vector form as follows
\begin{equation}
 \Omega' =f(\Omega),
\end{equation}
where $f$ can be extracted from(\ref{omemgaeq}). The critical
points, $\Omega^\ast$, namely the points at which the system will
stay provided it is initially at there, are given by the condition
\begin{equation}
 f(\Omega^{\ast})=0.\label{f}
\end{equation}
Their dynamical character is determined by the eigenvalues of the
matrix
\begin{equation}
 \frac{\partial f}{\partial\Omega}|_{\Omega=\Omega^{\ast}}.
\end{equation}
 If the real part of
the eigenvalues of a critical point is not zero, the point is said
to be {\em hyperbolic} \cite{CS}. In this case, the dynamical
character of the critical point is determined by the sign of the
real part of the eigenvalues:  If all of them are positive, the
point is said to be a {\em repeller}, because arbitrarily small
deviations from this point will move the system away from this
state.  If all of them are negative the point is called an {\em
attractor} because if we move the system slightly from this point
in an arbitrary way, it will return to it. Otherwise, we say the
critical point is a {\em saddle} point. \\We construct our models
as follows:
\\
\\
 \noindent {\bf (1)} \ The model $k_{\epsilon}$ , or
$(Z, \tilde{\Omega}_{k}, \tilde{\Omega}_{A}, \tilde{\Omega}_{C}) =
(\epsilon, 1, 0, 0)$, where $\epsilon\equiv \mbox{sgn}(Z)$. We
have
\begin{equation}
q=2-\frac{2}{Z^2}=0\, ,
\end{equation}
and the eigenvalues are
\begin{equation}
(\lambda_{Z}, \lambda_{k}, \lambda_{A}, \lambda_{C})=(0, 0, -4Z,
-12Z)=-\epsilon(0, 0, 4, 12)
\end{equation}

 \noindent {\bf (2)} \ The model $A_{\epsilon}$ , or
$(Z, \tilde{\Omega}_{k}, \tilde{\Omega}_{A}, \tilde{\Omega}_{C}) =
(\epsilon, 0, 1, 0)$. We have
\begin{equation}
q=2-\frac{1}{Z^2}=1\, ,
\end{equation}
and the eigenvalues are
\begin{equation}
(\lambda_{Z}, \lambda_{k}, \lambda_{A}, \lambda_{C})=(2Z, 2Z, 0,
-4Z)=\epsilon(2, 2, 0, -4)
\end{equation}

 \noindent {\bf (3)} \ The model $C_{\epsilon}$ , or
$(Z, \tilde{\Omega}_{k}, \tilde{\Omega}_{A}, \tilde{\Omega}_{C}) =
(\epsilon, 0, 0, 1)$. We have
\begin{equation}
q=2+\frac{3}{Z^2}=5\, ,
\end{equation}
and the eigenvalues are
\begin{equation}
(\lambda_{Z}, \lambda_{k}, \lambda_{A}, \lambda_{C})=(10Z, 10Z,
8Z, 4Z)=\epsilon(10, 10, 8, 4)
\end{equation}
\noindent {\bf (4)} The model $O$, or
$$
(Z, \tilde{\Omega}_{k},
\tilde{\Omega}_{A}, \tilde{\Omega}_{C}) = (0,
\tilde{\Omega}_{k}^{*}, \tilde{\Omega}_{A}^{*},
\tilde{\Omega}_{C}^{*})
$$
where $\tilde{\Omega}_{k}^{*}, \tilde{\Omega}_{A}^{*}$ and
$\tilde{\Omega}_{C}^{*}$ are constants satisfying (\ref{const1}) and
(\ref{f}). The eigenvalues of these points can be obtained in a
straightforward manner and show a saddle point, which has not been
included here. Hence, $O$ represents a set of infinite saddle points
whose line element is that of an open universe ($k=-1$) with $H=0$.

The dynamical system (\ref{omemgaeq}) has three
hyperbolic critical
points as follows: \\
\\
i) The model $k \:(k=-1)$,
$$
A=B=C=0, \:\: a(t)=t,
$$
with the critical point of an {\it attractor} type.\\
\\
ii) The model $A$,
$$
k=B=C=0, \:\: a(t)=(A)^{1/4}\sqrt{2t},
$$
with the critical point of a {\it saddle point} type. \\
iii) The model $C$,
$$
k=A=B=0, \:\: a(t)=(4\sqrt{C} t)^{1/4},
$$
with the critical point of a {\it repller}  type. \\
iv) The model $O$,
$$
Z=0, \tilde{\Omega}_{k}^*, \tilde{\Omega}_{A}^*,
\tilde{\Omega}_{C}^*,
$$
with the critical point of a {\it saddle point} type.

The four models have been depicted within the compact state space,
in Fig.(1). There are just trajectories on the planes, which are
invariant submanifolds of the state space.

\section{Conclusion}

In this work, we have studied bouncing cosmological solutions and
their stability with respect to homogeneous and isotropic
perturbations in a braneworld theory in which the bulk is given by a
five-dimensional charged AdS black hole spacetime with matter field
confined in a D3 brane. The effects of the charge of
five-dimensional black hole in the bulk has been considered. By
including this quantity in the analysis we have obtained four models
with the critical points of an {\it attractor}, a couple  {\it
saddle point} and a {\it repeller}, respectively, and constructed
the complete state space for these cosmological models.

\section*{Acknowledgment}

This work has been supported by Research Institute for Astronomy
and Astrophysics of Maragha.

\newpage
{\large {\bf Figure Captions}} \vspace{10mm}
\\
Figure 1. State space for the bouncing braneworld models. The
points $k_+, A_+$ and $C_+$ describe the critical points of an
{\it attractor}, a {\it saddle point} and a {\it repeller},
respectively. The points $k_-, A_-$ and $C_-$ describe the
critical points of a {\it repeller}, a {\it saddle point} and an
{\it attractor}, respectively. The points $O$ represent a set of
infinite saddle points. Only trajectories on the invariant planes,
$( \tilde{\Omega}_k=0, \tilde{\Omega}_A=0,$ and
$\tilde{\Omega}_C=0 )$ which outline the whole dynamics, are
drawn.
\end{document}